\pdfoutput=1
\documentclass[usenatbib]{tjaa}
\usepackage{graphicx}
\usepackage{subcaption}
\usepackage{longtable}
\usepackage{blindtext, alltt}
\usepackage{lipsum}
\usepackage{fancyhdr}
\usepackage{textcomp} 
\usepackage{amsmath}
\usepackage{xcolor}
\usepackage[authoryear]{natbib}
\usepackage{hyperref}  

\makeatletter
\newcommand{\@copyname}{} 
\newcommand{\@journal}{}  
\def\f@nch@olf{\copyright\ \pubyear\ \@copyname\strut}
\def\@thefoot{\small \hfil \@journal}
\makeatother

\newsavebox\verbbox

\title[The Femininomenon of Inequality]{The Femininomenon of Inequality: A Data-Driven Analysis and Cluster Profiling in Indonesia}

\author[J. S. Muthmaina]{%
Jessica Syafaq Muthmaina\autid{1}{0000-0000-0000-0000}\\
\adrid{1}Department of Physics, Faculty of Mathematics and Natural Sciences, Universitas Gadjah Mada%
}

\renewcommand{\pubyear}{2024}
\renewcommand{\volume}{0}
\renewcommand{\issue}{0}

\begin{document}
\label{firstpage}
\pagerange{\pageref{firstpage}--\pageref{lastpage}}
\maketitle{M00-0000}

\begin{abstract}
This study addresses the persistent challenges of Workplace Gender Equality (WGE) in Indonesia, examining regional disparities in gender empowerment and inequality through the Gender Empowerment Index (IDG) and Gender Inequality Index (IKG). Despite Indonesia's economic growth and incremental progress in gender equality, as indicated by improvements in the IDG and IKG scores from 2018 to 2023, substantial regional differences remain. Utilizing k-means clustering, the study identifies two distinct clusters of regions with contrasting gender profiles. Cluster 0 includes regions like DKI Jakarta and Central Java, characterized by higher gender empowerment and lower inequality, while Cluster 1 comprises areas such as Papua and North Maluku, where gender disparities are more pronounced. The analysis reveals that local socio-economic conditions and governance frameworks play a critical role in shaping regional gender dynamics. Correlation analyses further demonstrate that higher empowerment is generally associated with lower inequality and greater female representation in professional roles. These findings underscore the importance of targeted, region-specific interventions to promote WGE, addressing both structural and cultural barriers. The insights provided by this study aim to guide policymakers in developing tailored strategies to foster gender equality and enhance women’s participation in the workforce across Indonesia’s diverse regions.
\end{abstract}


\section{Introduction}
Workplace Gender Equality (WGE) remains a significant challenge in Indonesia. According to the World Bank’s 2020 \textit{Women, Business, and the Law} (WBL) Index, Indonesia ranks 144th out of 190 countries with a score of 64.4, indicating considerable gaps in legal protections for women. This low ranking suggests inadequate provisions to ensure equal pay, prevent workplace discrimination, and safeguard against sexual harassment, highlighting an urgent need for strengthened regulatory measures to promote gender equality in the workforce \citep{worldbank2020}.

The economic advantages of gender diversity in leadership are well-documented by the International Labour Organization (ILO), which links it to enhanced productivity, profitability, and innovation. Despite Indonesia’s status as Southeast Asia's largest economy, female representation in managerial roles remains low, with women concentrated in administrative positions, while men dominate decision-making roles. Addressing WGE in Indonesia could not only reduce these disparities but also foster career growth opportunities for women, thereby contributing to broader economic development \citep{ilo2020}.

Progress in gender equality is evident in recent data from Badan Pusat Statistik (BPS), which reports that Indonesia’s Gender Inequality Index (IKG) fell to 0.447 in 2023 due to improvements across multiple dimensions, particularly in various provinces \citep{bps2024}. Likewise, the Gender Empowerment Index (IDG) rose to 76.90 in 2023, indicating greater female representation in economic and political roles, including managerial and parliamentary positions \citep{bps2024_idg}. Further, the United Nations Development Programme (UNDP) reports a Gender Development Index (GDI) of 91.85 in 2023, along with an increase in the Women’s Financial Inclusion Index to 83.88, reflecting a positive shift towards financial independence and inclusion for women \citep{undp2023}.

However, substantial challenges persist. According to the World Economic Forum’s 2023 \textit{Global Gender Gap Report}, Indonesia ranks 87th out of 146 countries with a score of 0.697, showing only moderate progress in economic participation, educational attainment, and political empowerment. While women are approaching parity in technical roles, their representation in senior positions has declined, and wage gaps remain, with women earning only 51.9 cents for every dollar earned by men \citep{wef2023}.

This study aims to address these persistent disparities by analyzing regional gender inequalities across Indonesia using the Gender Empowerment Index (IDG) and Gender Inequality Index (IKG). Employing clustering techniques, this research identifies regional patterns and examines the drivers of gender gaps across provinces, ultimately providing data-driven insights to guide policymakers in developing targeted, region-specific interventions to advance workplace gender equality and sustain progress in gender equity nationwide.

\section{Methodology}
This research utilized Python for data analysis and modeling, incorporating various libraries to support data processing and visualization. The primary datasets included the Gender Empowerment Index (IDG), Gender Inequality Index (IKG), and the percentage of female professionals across Indonesia’s regions, covering the years 2018 to 2023. These data were sourced from the official BPS website, followed by cleaning and preprocessing to ensure consistency and accuracy in the analysis.

The study employed k-means clustering to classify Indonesian regions based on IDG and IKG scores, aiming to reveal underlying patterns of gender disparity and empowerment across different areas. The clustering approach allowed us to group regions with similar gender equality profiles, facilitating a comparative analysis. After preliminary tests, we selected \( k=2 \) as the optimal number of clusters to categorize regions with relatively high or low levels of gender empowerment and inequality. To evaluate the clustering quality, we used metrics such as the silhouette score, which assess the coherence and separability of the clusters.

\section{Results and Discussion}
\subsection{Time Analysis}
The analysis reveals distinct regional trends in gender empowerment, gender inequality, and female professional participation across Indonesia from 2018 to 2023.

\begin{figure}
    \centering
    \includegraphics[width=1\linewidth]{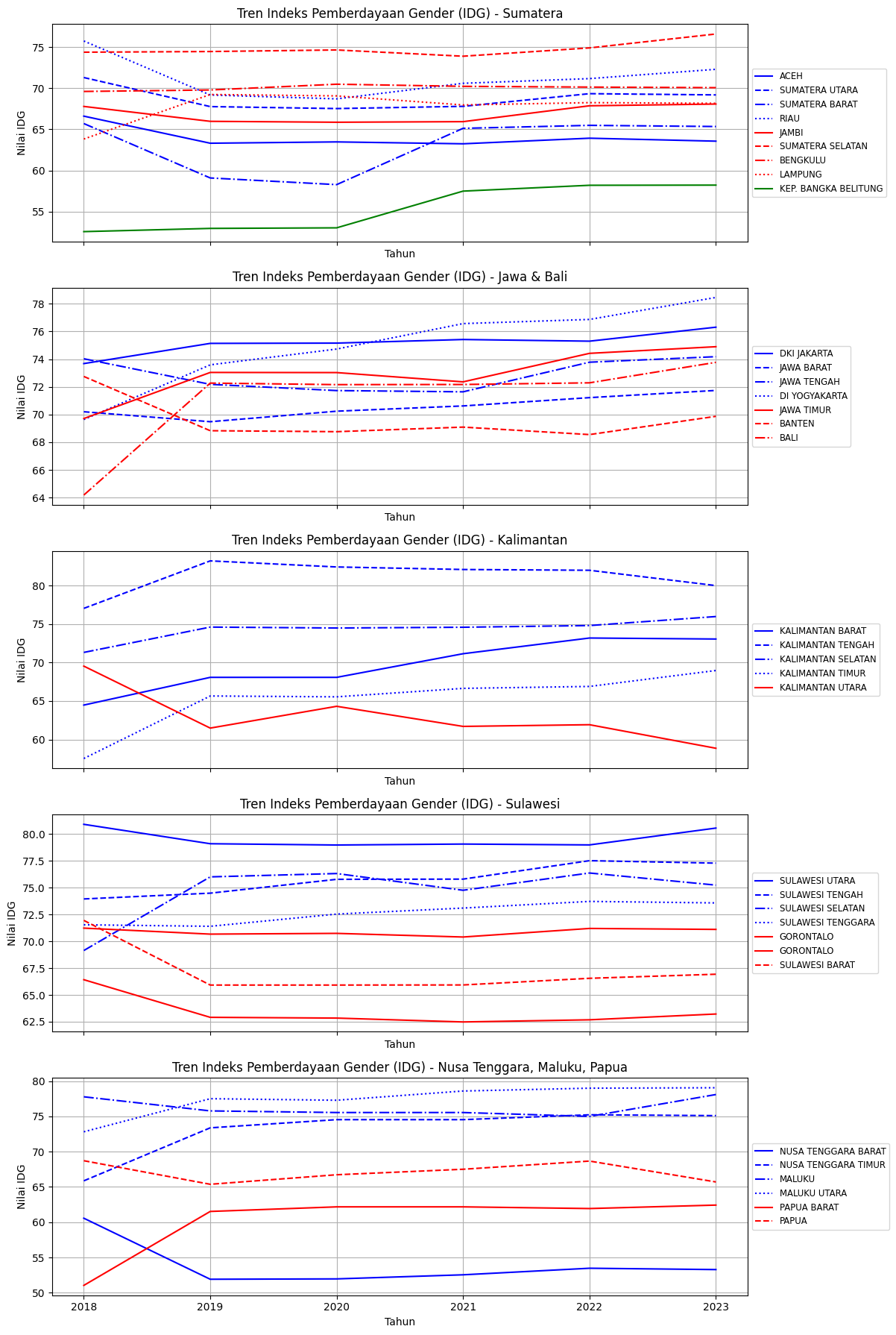}
    \caption{The Gender Empowerment Index (IDG) Trends}
    \label{fig:TS_IDG}
\end{figure}

The Gender Empowerment Index (IDG) demonstrates consistent improvement in several regions, notably in Java and Bali, where empowerment levels for women have steadily increased. This upward trend suggests effective policy implementation and social factors that support gender empowerment. In contrast, regions such as Kalimantan and Sulawesi exhibit more variability, with empowerment levels fluctuating over time. This variation indicates that while national policies contribute to overall progress, local socio-economic conditions play a critical role in shaping the degree and sustainability of gender empowerment at the regional level.

\begin{figure}
    \centering
    \includegraphics[width=1\linewidth]{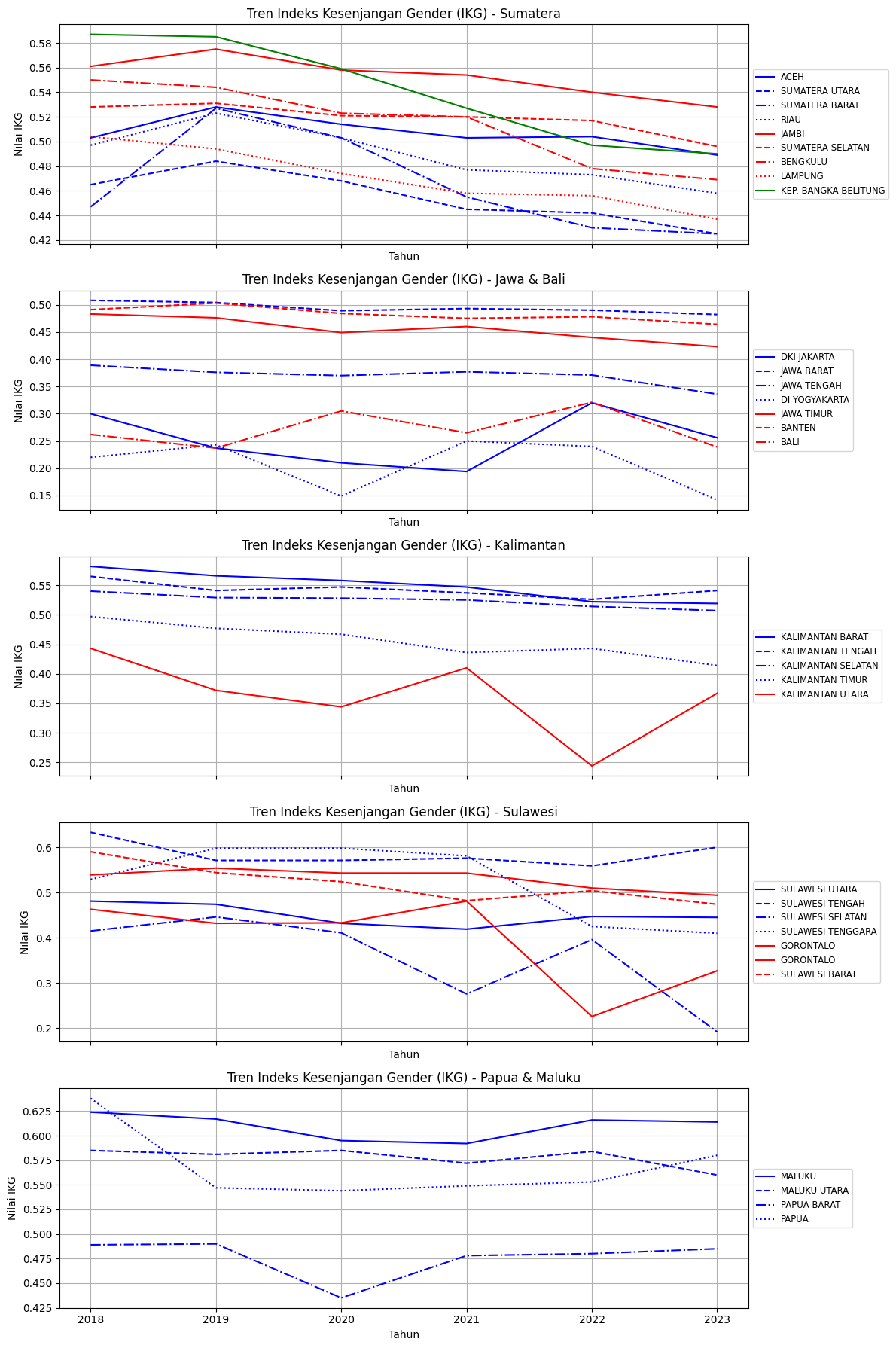}
    \caption{The Gender Inequality Index (IKG) Trends}
    \label{fig:TS_IKG}
\end{figure}

Similarly, the Gender Inequality Index (IKG) trends, as shown in Figure \ref{fig:TS_IKG}, underscore ongoing gender disparities across Indonesia. In regions like Sumatra and Kalimantan, IKG values show fluctuations with only slight improvements, reflecting persistent barriers to achieving gender equality. Meanwhile, Java has achieved more stable, though gradual, declines in inequality levels, likely supported by more robust institutional frameworks and resources dedicated to promoting gender equality. These findings suggest that structural challenges unique to each region may either aid or hinder progress, highlighting the need for tailored approaches in addressing gender disparities.

\begin{figure}
    \centering
    \includegraphics[width=1\linewidth]{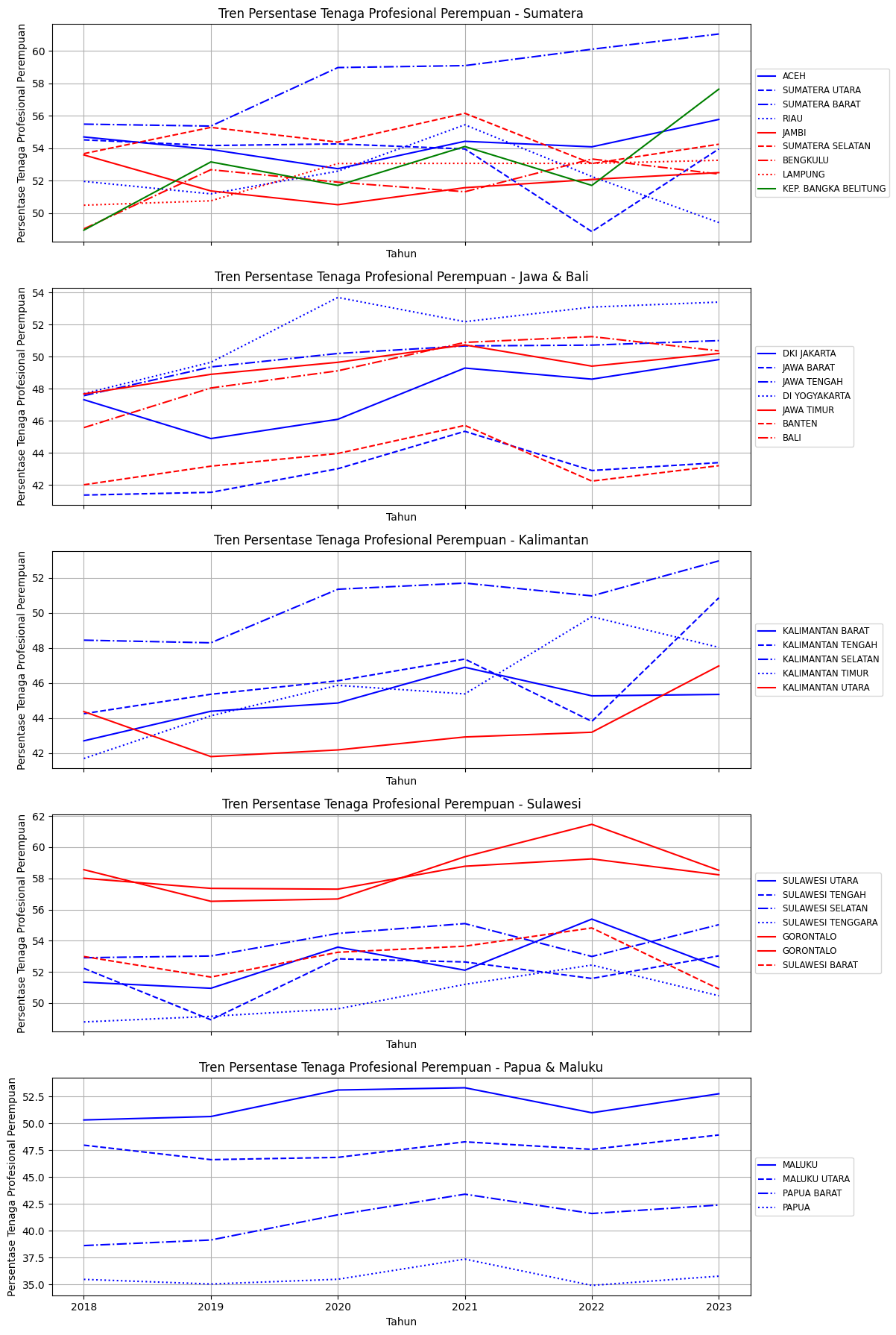}
    \caption{Trend in Percentage of Female Professional Workforce}
    \label{fig:TS_Tenaga_Professional}
\end{figure}

The analysis of female professional participation reveals a general upward trend, with a notable increase in the proportion of women in professional roles in many regions, as seen in Figure \ref{fig:TS_Tenaga_Professional}. However, significant regional disparities remain, as evidenced by limited progress in Papua and Maluku. By contrast, Sulawesi demonstrates substantial advancements in female professional representation, underscoring the influence of local policies and socio-economic conditions in encouraging women's participation in professional fields. These regional differences highlight the critical role of targeted interventions to foster gender equality and support women’s professional growth, particularly in regions where progress has been slower.

\subsection{K-Means Clustering}
In this study, k-means clustering was applied to identify patterns of similarity between Indonesian provinces based on their Gender Empowerment Index (IDG) and Gender Inequality Index (IKG) scores. K-means was selected for its simplicity and effectiveness in grouping data points with similar characteristics. After evaluating various values of \( k \) using the silhouette score—a metric that assesses clustering quality by measuring cohesion within clusters and separation between them—the optimal \( k \) was determined to be 2. This choice of \( k=2 \) provided the highest silhouette score, indicating well-defined, cohesive clusters that enhance interpretability.

\subsubsection{Silhouette Score and Clustering of IKG vs IDG}
\begin{figure}
    \centering
    \begin{subfigure}[b]{0.45\linewidth}
        \centering
        \includegraphics[width=\linewidth]{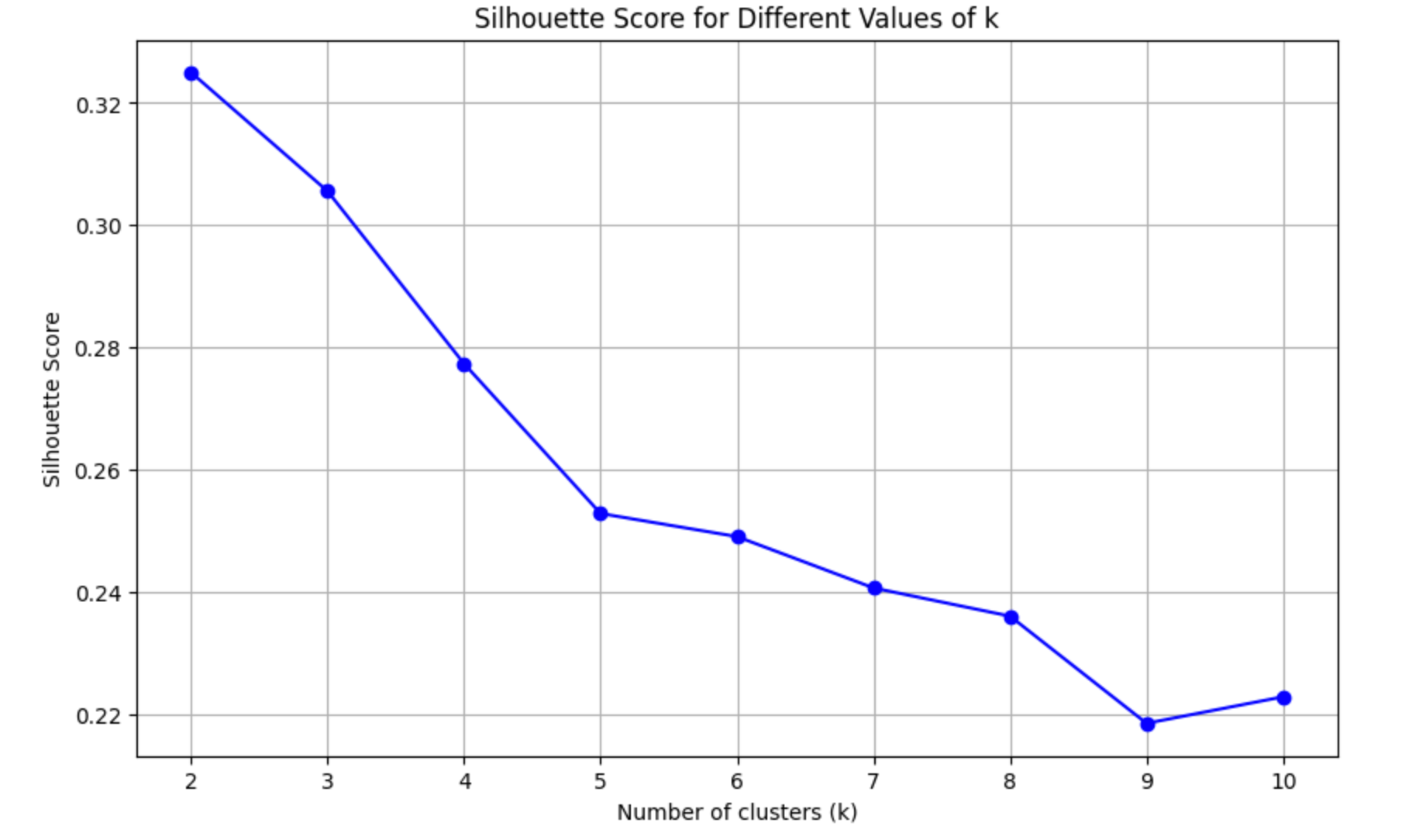}
        \caption{Silhouette Score}
        \label{fig:SS}
    \end{subfigure}
    \hfill
    \begin{subfigure}[b]{0.45\linewidth}
        \centering
        \includegraphics[width=\linewidth]{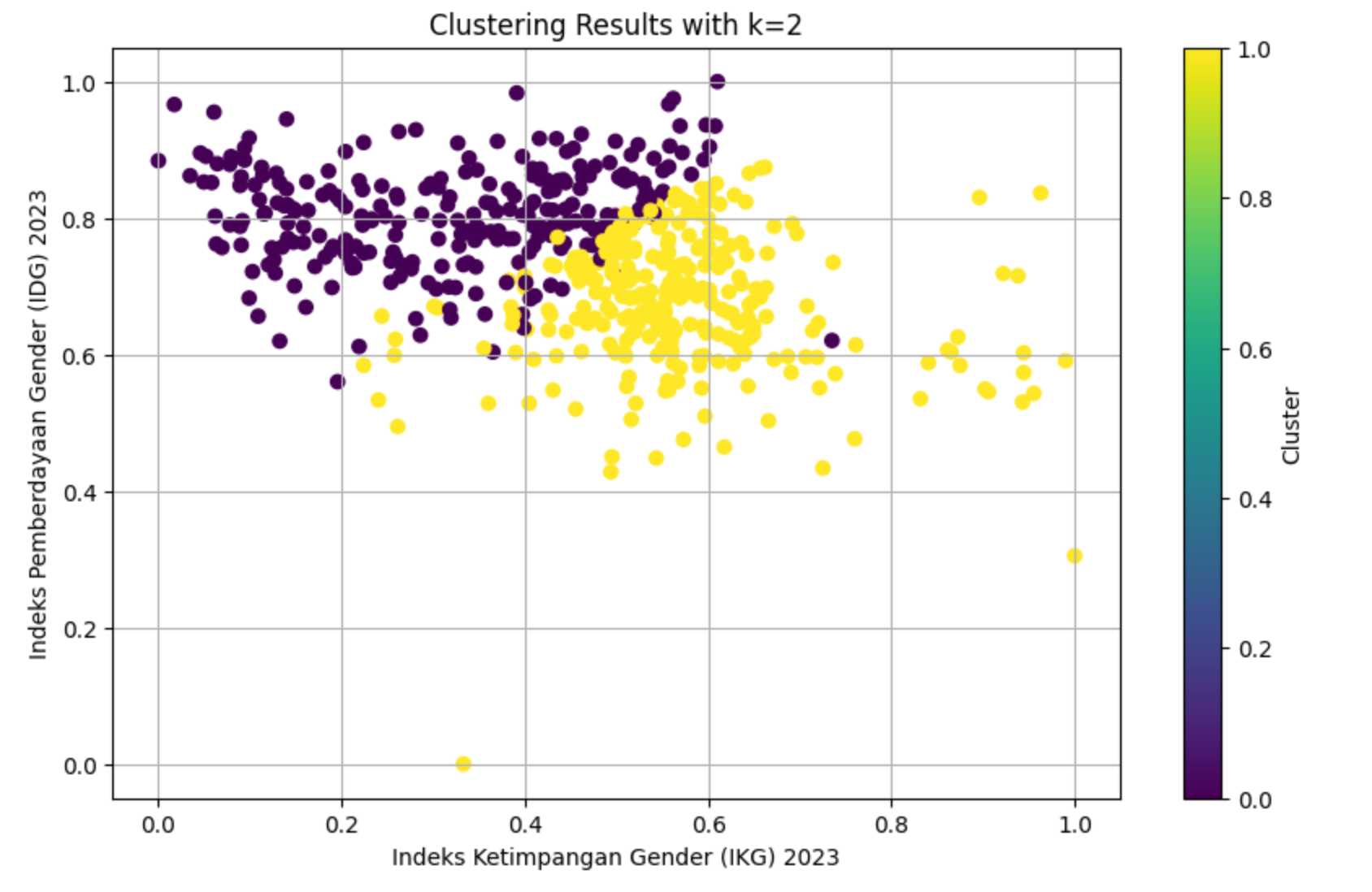}
        \caption{Clustering of IKG vs IDG}
        \label{fig:clustering}
    \end{subfigure}
    \caption{Comparison of Silhouette Score and Clustering of IKG vs IDG}
    \label{fig:side_by_side}
\end{figure}

Based on the clustering results using K-Means with \( k=2 \), there are two clusters, as shown in Figure \ref{fig:SS}. Cluster 0 comprises provinces with relatively higher IDG scores and lower IKG values, reflecting greater female empowerment and lower gender inequality, such as DKI Jakarta and Central Java. Conversely, Cluster 1 includes provinces with higher gender inequality and lower women’s empowerment, such as Papua and North Maluku. These findings suggest that regions in Cluster 1 may require prioritized interventions to support gender equality and empower women economically and socially.

\subsubsection{Distribution of IKG Based on Cluster}

Further profiling of the clusters reveals significant differences in gender dynamics. Cluster 0 provinces show a higher average IDG and lower average IKG, indicative of better female empowerment and reduced gender inequality. Cluster 1, on the other hand, displays a contrasting profile, with lower average IDG and higher IKG values, highlighting areas with more severe gender disparities. 

\begin{figure}
    \centering
    \begin{subfigure}[b]{0.45\linewidth}
        \centering
        \includegraphics[width=\linewidth]{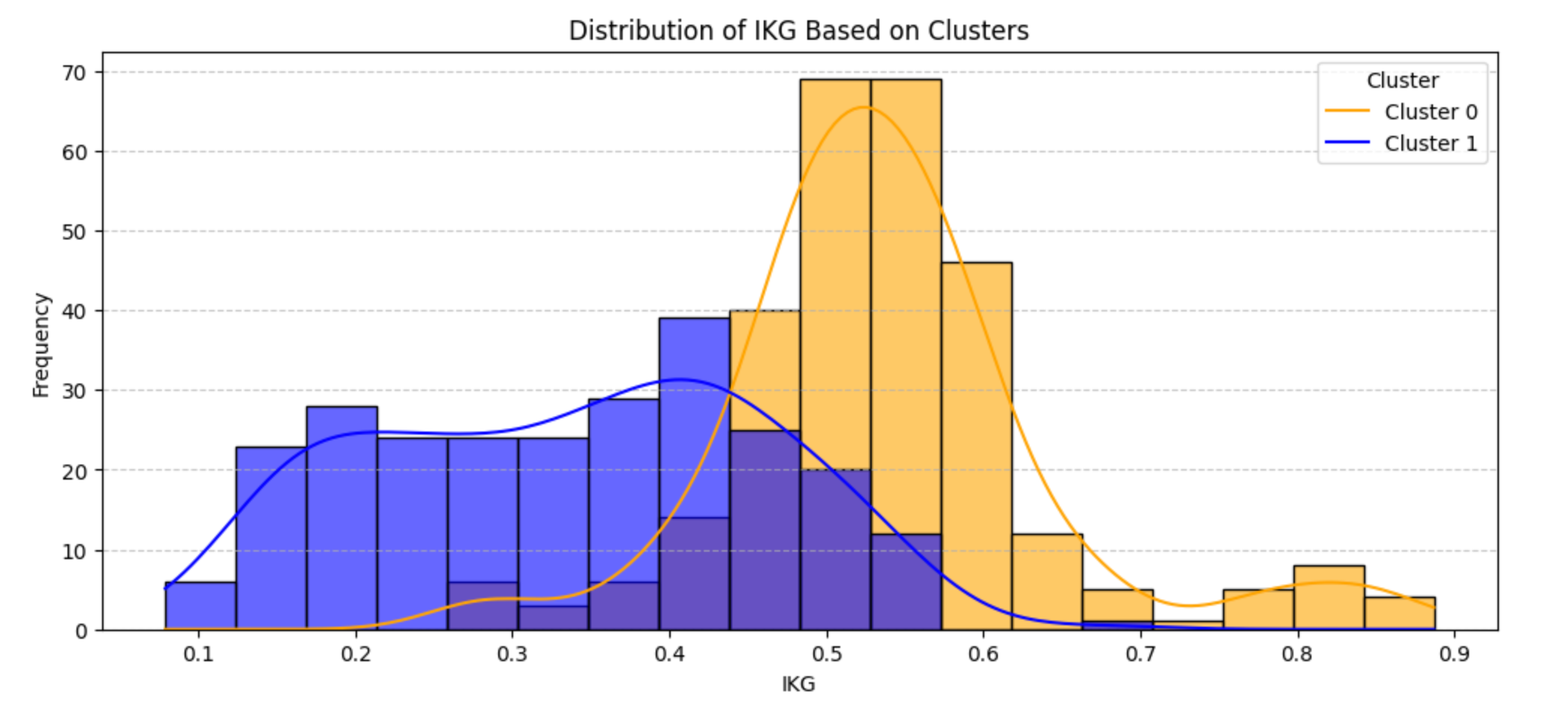}
        \caption{Distribution of IKG based on Clustering}
        \label{fig:IKG_clust}
    \end{subfigure}
    \hfill
    \begin{subfigure}[b]{0.45\linewidth}
        \centering
        \includegraphics[width=\linewidth]{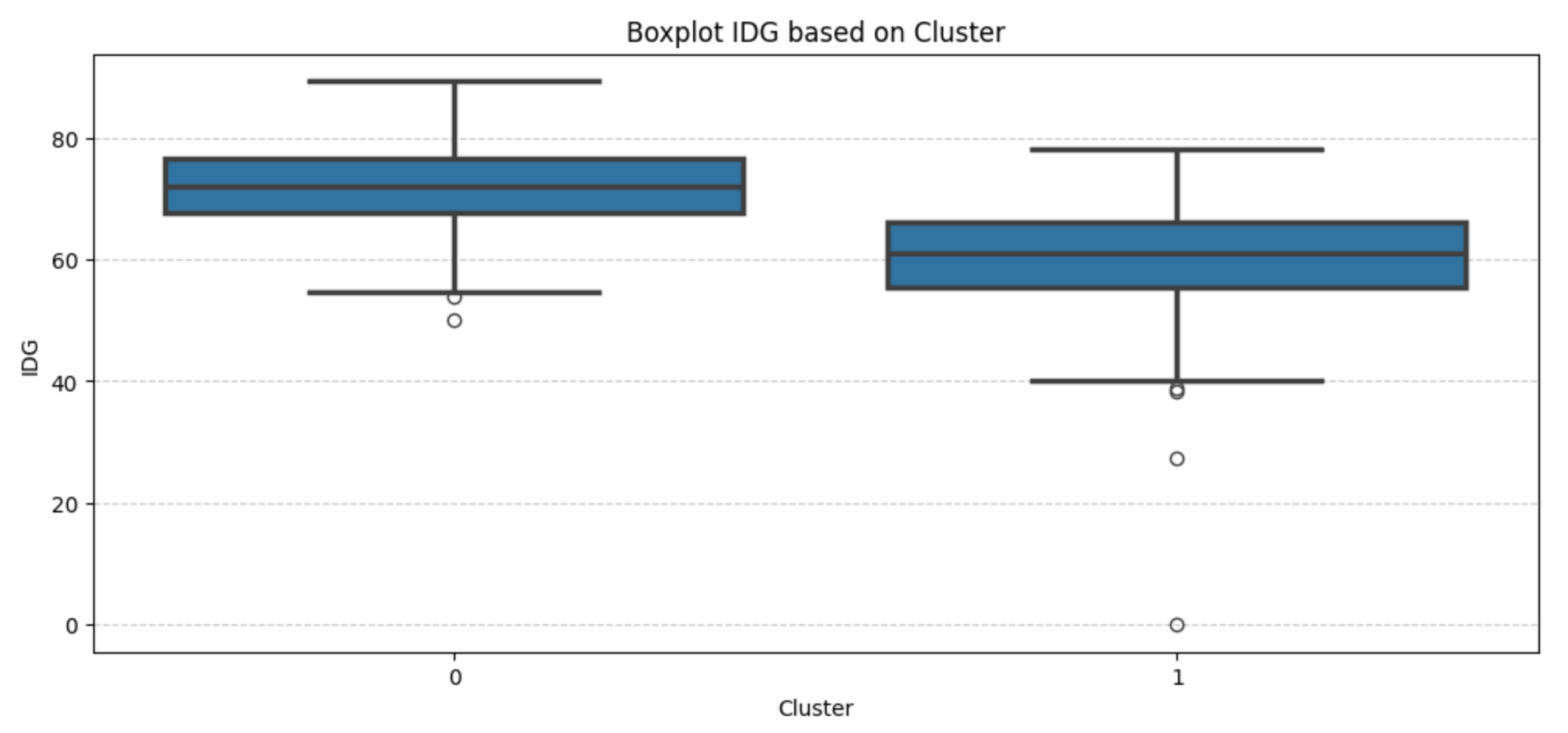}
        \caption{Boxplot IDG based on Cluster}
        \label{fig:IDG_clust}
    \end{subfigure}
    \caption{Visual representation of the clustering process, including IKG distribution and IDG boxplot based on clusters.}
    \label{fig:clustering_side_by_side}
\end{figure}

The visualization of IKG distribution in Figure \ref{fig:IKG_clust} further illustrates these differences between the two clusters. Cluster 0 shows a lower IKG distribution, reflecting lower levels of gender inequality, while Cluster 1 exhibits a higher distribution of IKG values, indicative of more pronounced gender disparities. These distinctions underscore the contrasting gender equality profiles of the two clusters, guiding targeted policy approaches tailored to the specific needs of each group.

The boxplot analysis of the Gender Empowerment Index (IDG) for each cluster in Figure \ref{fig:IDG_clust} reveals distinct differences in IDG values. Cluster 0 displays a higher median IDG, indicating better levels of female empowerment within these provinces, whereas Cluster 1 has a lower median IDG, suggesting relatively lower levels of female empowerment. These median disparities highlight the contrasting empowerment profiles of the clusters, providing insights into the varying degrees of gender empowerment across different regions.

\subsection{Heatmap between IDG, IKG, and Percentage of Female Professionals}

\begin{figure}
    \centering
    \includegraphics[width=1\linewidth]{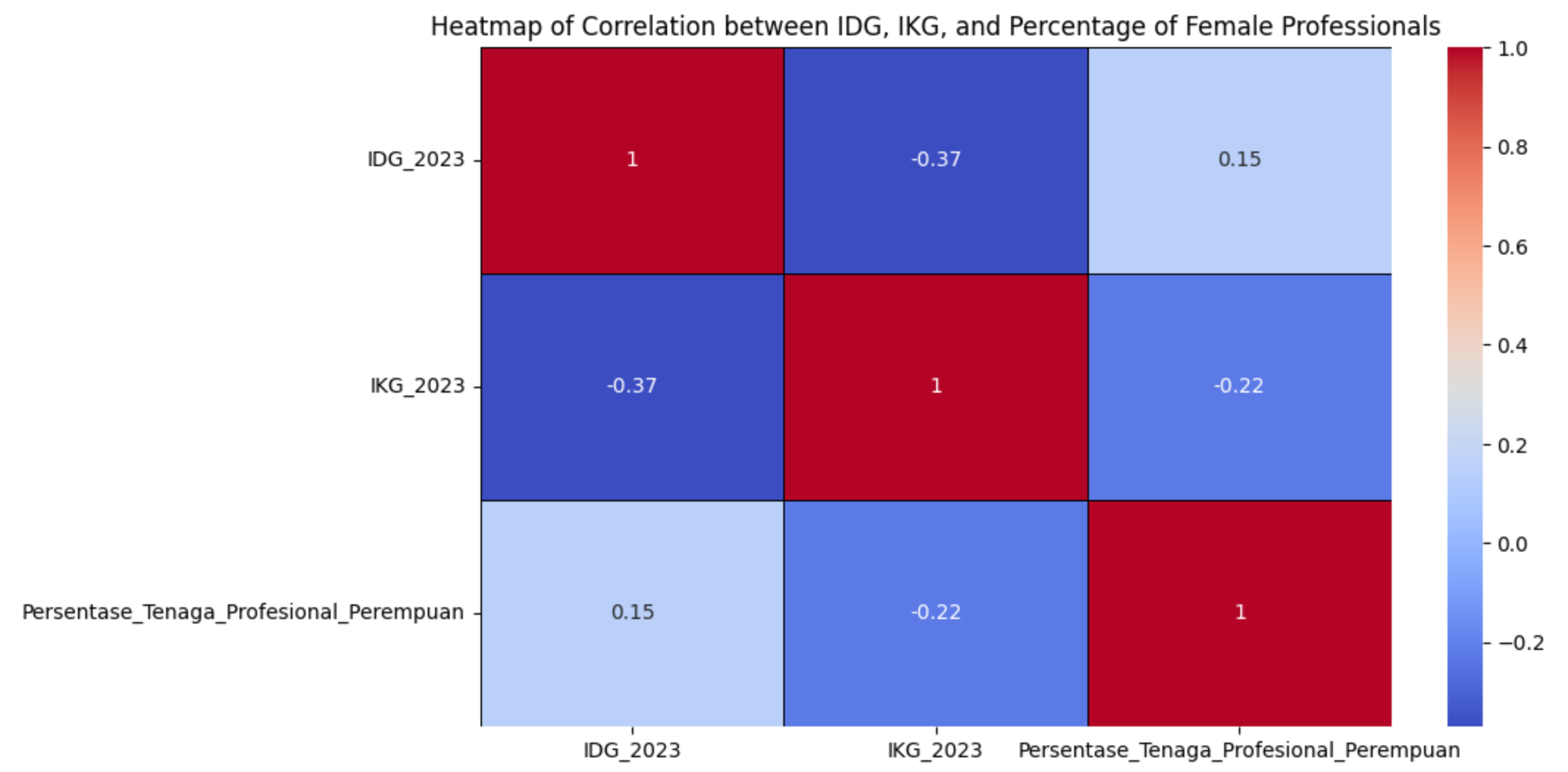}
    \caption{Heatmap between IDG, IKG, and Percentage of Female Professionals}
    \label{fig:heatmap}
\end{figure}

The heatmap in Figure \ref{fig:heatmap} visualizes correlations among the Gender Empowerment Index (IDG), Gender Inequality Index (IKG), and the percentage of female professionals for 2023. A moderate negative correlation between IDG and IKG (-0.37) suggests that as gender empowerment increases, gender inequality tends to decrease. Additionally, there is a weak positive correlation between IDG and the percentage of female professionals (0.15), indicating that higher gender empowerment is associated with a modest increase in women's professional representation. Conversely, the negative correlation between IKG and the percentage of female professionals (-0.22) implies that lower gender inequality may facilitate greater female participation in professional occupations. These correlations underscore the interconnectedness of gender empowerment, inequality, and women’s professional involvement, highlighting the importance of policies that simultaneously address these dimensions to foster an inclusive professional landscape for women.

\subsection{Pairplot of Relationship Between Indicators Based on Cluster}

\begin{figure}
    \centering
    \includegraphics[width=1\linewidth]{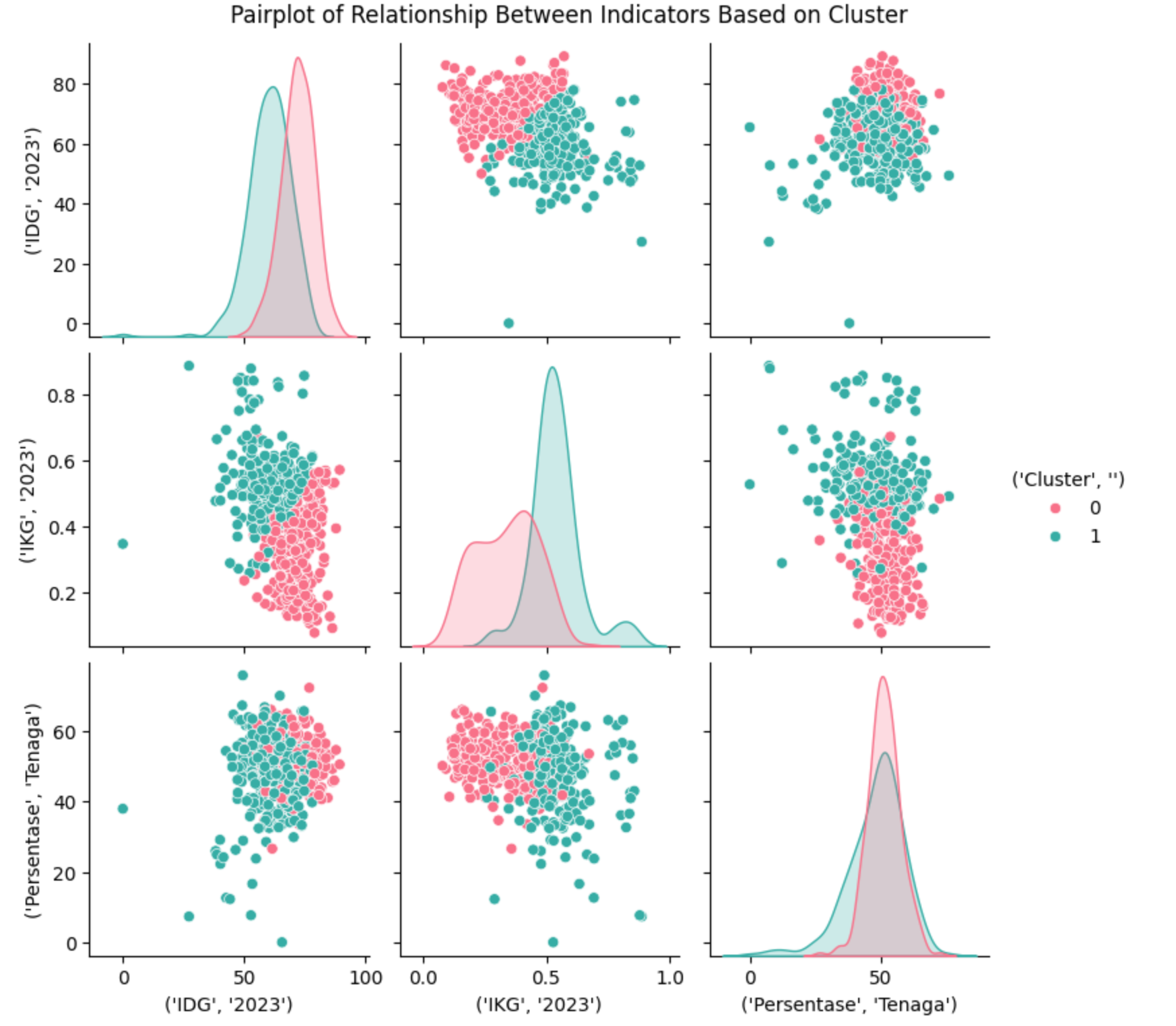}
    \caption{Pairplot of Relationship Between Indicators Based on Cluster}
    \label{fig:pairplot}
\end{figure}

The pairplot in Figure \ref{fig:pairplot} offers a segmented view of the relationships among IDG, IKG, and the percentage of female professionals, with each point representing data from either Cluster 0 or Cluster 1. Cluster 0, shown in red, generally has lower IKG values and higher IDG values, indicating stronger gender empowerment and lower inequality. In contrast, Cluster 1, depicted in teal, typically exhibits higher IKG values and lower IDG scores, suggesting regions with more pronounced gender inequality and reduced female empowerment. The distribution and scatter plots reveal distinct patterns within each cluster, reflecting structural differences in gender dynamics across the regions.

These visualizations clarify the contrasting characteristics of the clusters, providing deeper insight into the varied gender equality profiles across Indonesian regions. This differentiation highlights how regions in Cluster 1 may benefit from targeted interventions aimed at reducing gender inequality and enhancing professional opportunities for women. Together, the heatmap and pairplot analyses offer a comprehensive view of the factors influencing gender dynamics, thereby guiding policymakers in crafting region-specific strategies to promote gender empowerment and workforce equality.

\section{Conclusions}
This study provides a comprehensive analysis of gender inequality across Indonesia, highlighting distinct regional variations in gender empowerment, inequality, and professional participation among women from 2018 to 2023. The findings reveal that while Indonesia has made strides in improving gender-related indices nationally, significant disparities remain at the regional level, influenced by socio-economic conditions, local governance, and access to resources. Through the use of the Gender Empowerment Index (IDG) and the Gender Inequality Index (IKG), this study identifies patterns and trends that differentiate regions, suggesting that some areas have achieved more stable and consistent progress, while others continue to face barriers that hinder gender equality.

The clustering analysis, which divided the regions into two distinct clusters, underscores the uneven nature of gender equality progress across Indonesia. Cluster 0 includes regions such as DKI Jakarta and Central Java, where higher IDG values and lower IKG scores reflect relatively better outcomes for women in terms of empowerment and reduced inequality. These regions benefit from more robust economic development, greater access to resources, and potentially stronger institutional support for gender equality initiatives. By contrast, Cluster 1 comprises regions like Papua and North Maluku, where lower IDG values and higher IKG scores point to higher levels of gender inequality and less empowerment for women. This disparity highlights the persistent socio-economic and structural challenges in these regions, including limited access to education, healthcare, and formal employment opportunities, as well as entrenched cultural norms that may restrict women’s participation in the workforce and leadership roles.

In terms of professional representation, the analysis found a general upward trend in the proportion of women in professional roles across many regions, but significant disparities remain. For instance, while Sulawesi shows substantial advancements in female professional participation, other regions like Papua and Maluku exhibit minimal progress. This variation underscores the role of local policies and socio-economic conditions in influencing women’s professional involvement. Collectively, these insights provide a nuanced view of how regional disparities in gender dynamics impact women’s opportunities, suggesting that policies need to be tailored to the specific needs and conditions of each region.

\section{Future Work}
Future research should expand on this analysis by incorporating additional socio-economic and cultural variables to deepen our understanding of the factors influencing gender disparities. Variables such as access to education, healthcare quality, and regional economic growth could provide a more comprehensive view of the underlying drivers of gender inequality. Additionally, longitudinal studies that track the impact of newly implemented gender policies over time would allow for a more dynamic assessment of policy effectiveness. Advanced machine learning techniques and mixed-method approaches, such as integrating qualitative data from stakeholder interviews, could further refine the analysis and add valuable contextual insights. This expanded approach would support policymakers in designing more nuanced, evidence-based strategies that are both responsive to regional conditions and effective in promoting sustained progress in workplace gender equality across Indonesia.

\bibliographystyle{plainnat}
\bibliography{refs}

\label{lastpage}
\end{document}